\documentclass[sigconf]{acmart}

\AtBeginDocument{%
  \providecommand\BibTeX{{%
    \normalfont B\kern-0.5em{\scshape i\kern-0.25em b}\kern-0.8em\TeX}}}

\copyrightyear{2019}
\acmYear{2019}
\acmConference[NANOCOM '19]{The Sixth Annual ACM International Conference on
  Nanoscale Computing and Communication}{September 25--27, 2019}{Dublin,
  Ireland}
\acmBooktitle{The Sixth Annual ACM International Conference on Nanoscale
  Computing and Communication (NANOCOM '19), September 25--27, 2019, Dublin,
  Ireland}
\acmPrice{15.00}
\acmDOI{10.1145/3345312.3345497}
\acmISBN{978-1-4503-6897-1/19/09}

\newcommand{\ket}[1]{| #1 \rangle}
\newcommand{\Complex}{\mathbb{C}}

\begin{document}

\title{Towards Large-Scale Quantum Networks}

\author{Wojciech Kozlowski}
\email{w.kozlowski@tudelft.nl}
\affiliation{%
  \institution{QuTech, Delft University of Technology}
  \streetaddress{Lorentzweg 1}
  \city{Delft}
  \country{Netherlands}
  \postcode{2628 CJ}
}

\author{Stephanie Wehner}
\email{s.d.c.wehner@tudelft.nl}
\affiliation{%
  \institution{QuTech, Delft University of Technology}
  \streetaddress{Lorentzweg 1}
  \city{Delft}
  \country{Netherlands}
  \postcode{2628 CJ}
}

\begin{abstract}
  The vision of a quantum internet is to fundamentally enhance Internet
  technology by enabling quantum communication between any two points on Earth.
  While the first realisations of small scale quantum networks are expected in
  the near future, scaling such networks presents immense challenges to
  physics, computer science and engineering.  Here, we provide a gentle
  introduction to quantum networking targeted at computer scientists, and
  survey the state of the art. We proceed to discuss key challenges for
  computer science in order to make such networks a reality.
\end{abstract}



\maketitle

\section{Introduction}
\begin{figure}[h]
  \centering
  \includegraphics[width=\linewidth]{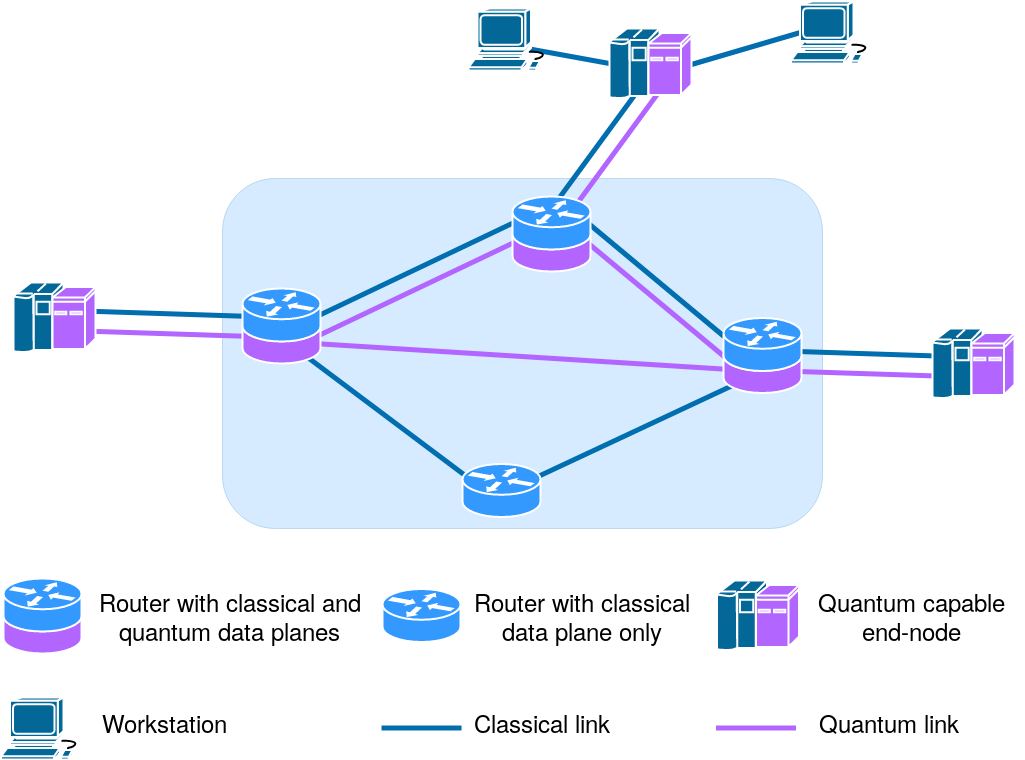}
  \caption{Quantum networks will be embedded within classical networks and use
    existing infrastructure to send and receive control messages. This can be
    achieved by adding a quantum data plane to existing networks. Note that the
    quantum and classical links do not have to coincide.}\label{fig:network}
  \Description{Quantum and classical networks.}
\end{figure}

The objective of quantum networks is to fundamentally enhance communication
technology by allowing the transmission and manipulation of quantum bits
(qubits) between remote locations. Such networks will be embedded within
classical networks as shown in Fig.~\ref{fig:network} and applications will
have access to both quantum and classical channels. Quantum networks will be
used to execute protocols that have no classical counterpart or are more
efficient than what is possible classically. The range of possible quantum
applications will depend on the development stage of the underlying hardware
hardware~\cite{wehner2018quantum}. This new networking paradigm has already
opened up a range of new applications, which are provably impossible to realise
using classical communication over the internet that we have today. Quantum key
distribution (QKD)~\cite{bennett1984quantum, ekert1991quantum} to ensure secure
communication is the most famous example as it is also the only application
that is ready for commercialisation and is undergoing standardisation. Whilst
QKD will be the main focus for most near-term quantum networks, many other
applications have already been put forward, with many more to be expected when
such networks become widespread such as secure quantum computing in the
cloud~\cite{broadbent2009universal, fitzsimons2017unconditionally}, clock
synchronisation~\cite{komar2014quantum}, and sensor
networks~\cite{giovannetti2004quantum, gottesman2012longer}.

Using features of quantum mechanics as the underlying physical mechanism for
communication opens up many new possibilities, but also introduces considerable
new design challenges. Some of these design challenges are due to fundamental
differences between quantum and classical information, while others arise from
technological limitations in engineering large-scale quantum systems. The first
fundamental difference that quantum communication brings with it is the
no-cloning theorem~\cite{nielsen2000quantum}. That is, arbitrary quantum data
cannot be copied without destroying the original version. This means that it is
impossible to use the same solutions that worked for classical networks which
rely heavily on the ability to read and copy data for the purposes of
retransmission and signal amplification. These limitations make transmitting
qubits over long distances particularly challenging. The second fundamental
difference arises due to a phenomenon called quantum entanglement. Entanglement
is a special state of two or more qubits, that can in principle persist even if
they are separated by arbitrary geographical distances and it is the key
ingredient that enables long distance quantum communication. This property
exists at the physical level and it requires that the location and state of its
constituent qubits be known at all times. This is in contrast to classical
communication, where signals at the physical layer typically proceed from the
sender to the receiver and no state or notion of a connection exists. This
introduces new demands for the control of such networks, as quantum data is
inherently delocalised across multiple devices.

Design considerations that come from technological and not just fundamental
limitations form an integral part of quantum network development and a key
issue when considering realistic deployments.  The technological challenges are
immense, and include --- for example --- storing qubits for a long time or
manipulating a large number of qubits simultaneously.

The remainder of this paper is structured as follows:
Section~\ref{sec:state-of-the-art} briefly surveys the current state of the art
of quantum networked technologies and in
Section~\ref{sec:qubits-and-entanglement} we give a basic introduction to the
quantum physics of such networks. In
Sections~\ref{sec:elements-of-a-quantum-network}
and~\ref{sec:a-quantum-network-stack} we discuss the elements of a quantum
network and a possible network stack respectively. Future research challenges
are presented in Section~\ref{sec:challenges-and-requirements} and the paper is
concluded in~\ref{sec:conclusion}.

\section{State of the art}\label{sec:state-of-the-art}

At present, no large-scale quantum networks exist. At short distances
(\textasciitilde{}100~km in telecom fibre), devices that perform QKD are
commercially available~\cite{giovannetti2004quantum, inagaki2013entanglement,
  diamanti2016practical, fibresystems}. Early-stage demonstrations also achieve
longer distances in the lab using coiled fibre~\cite{boaron2018secure,
  stucki2009high, hiskett2006long, minder2019experimental, zhong2019proof,
  wang2019beating}, or through free space
communication~\cite{schmitt2007experimental, vallone2015experimental}. QKD
devices have been deployed in a variety of field tests and short-distance
networks~\cite{sasaki2011field, peev2009secoqc, stucki2011long, wang2014field}.

While no long-distance quantum networks exist, short distance segments have
been chained together classically to form so-called trusted repeater or trusted
node networks~\cite{salvail2010security, scarani2009security}. Such networks do
not allow the end to end transmission of qubits, or the generation of
entanglement and hence do not offer end-to-end security. They only enable
secure communication between two end-points, provided that all the intermediate
nodes are trusted. Such links of trusted nodes have been
realised~\cite{courtland2016china, sasaki2011field}, but require a high level
of physical security to protect the trusted nodes. Such devices only produce
short-lived (entanglement is not stored), short-distance entanglement and lack
any of the features needed to bridge longer distances.

Long-range quantum communication, as well as the realisations of networks with
functionalities more advanced than QKD, are presently still in their infancy.
Entanglement between distant sites (\textasciitilde{}1200~km) has been produced
using a satellite~\cite{yin2017satellite}. However, data rates
(\textasciitilde{}1~hz for 275~s per day) are still too low to produce a secret
key, and the entanglement is short-lived. The present record for producing
heralded entanglement between distant sites is 1.3~km in a solid state quantum
device (nitrogen-vacancy (NV) centres in diamond)~\cite{hensen2015loophole}.
Longer distances have been observed for nodes in the same
lab~\cite{yu2019entanglement}. Demonstrations of more complex applications such
as blind quantum computing~\cite{barz2012demonstration} and quantum
sensing~\cite{guo2019distributed} have also been realised in laboratory
conditions.

Going forward, we would like to improve early-stage quantum communication in
three directions. First, we would like to enable untrusted long-distance
communication. Second, we would also like to enable the execution of more
complex quantum network applications in order to take full advantage of our
ability to transmit qubits. And finally, we would like to improve accessibility
by allowing early stage access to such technology. The first realisation of
such a network, a four-node demonstration in the Netherlands, is scheduled to
be operational within the next 5--6 years. Much essential work is being done to
build quantum hardware to make this possible, which is covered at length in the
physics literature~\cite{sangouard2011quantum, munro2015inside,
  wehner2018quantum}.

\section{Qubits and Entanglement}\label{sec:qubits-and-entanglement}

This subsection will briefly introduce the basic concepts of quantum computing
and networking: qubits, quantum gates, and entanglement.  For additional
information see e.g.~\cite{nielsen2000quantum}.

\subsection{Qubits}

The differences between quantum computation and classical computation begin at
the bit-level. A classical computer operates on the binary alphabet $\{ 0, 1
\}$. Mathematically, a quantum bit, a qubit, exists over the same binary space,
but unlike the classical bit, it can exist in a so-called superposition of the
two possibilities:
\begin{equation}
  \ket{\Psi} = \alpha | 0 \rangle + \beta | 1 \rangle
\end{equation}
where $| X \rangle$ denotes a quantum state, here the binary $0$ and $1$, and
the coefficients $\alpha$ and $\beta$ are complex numbers called probability
amplitudes satisfying $|\alpha|^2 + |\beta|^2 = 1$.

Upon measurement\footnote{In the standard basis, given by
$\{\ket{0},\ket{1}\}$}, the qubit loses its superposition and irreversibly
collapses into one of the two basis states, either $| 0 \rangle$ or $| 1
\rangle$, and yields the corresponding value, $0$ or $1$, as the measurement
readout. The outcome of the measurement is not deterministic, and the
probability of measuring $0$ and collapsing the state to $| 0 \rangle$ is $|
\alpha |^2$ and similarly the probability of measuring $1$ and collapsing the
state to $|1\rangle$ is $| \beta |^2$. This randomness is not due to our
ignorance of the underlying mechanisms, but rather it is a fundamental feature
of a quantum mechanical system.

Many possible realisations of qubits exists. Key to all these representations,
is to find a realisation of the classical states $\ket{0}$ and $\ket{1}$,
together with a procedure to create arbitrary superpositions $\ket{\Psi}$
thereof. For quantum memories, and quantum computing devices, $\ket{0}$ and
$\ket{1}$ typically correspond to states of two different energies in either a
natural ``atomic system'' (e.g.\ ion traps~\cite{haffner2008quantum}, NV
centres in diamond~\cite{togan2010quantum}, neutral
atoms~\cite{briegel2000quantum} or atomic
ensembles~\cite{sangouard2011quantum}), or artificially designed nano-scale
systems (e.g.\ superconducting quantum
processors~\cite{clarke2008superconducting}). For transmission, usually
optically, qubits can be represented in a variety of ways: the two states
$\ket{0}$ and $\ket{1}$ can be encoded in the presence or absence of a
photon~\cite{humphreys2018deterministic, cabrillo1999creation}, a time-bin
encoding of early and late arrival~\cite{brendel1999pulsed}, or the horizontal
and vertical polarisation of photons~\cite{bennett1984quantum,
  mattle1996dense}.

\subsection{Multiple Qubits}

We can express the state of an $n$-qubit quantum state as
\begin{align}
\ket{\Psi} = \sum_{x \in {\{0,1\}}^n} \alpha_x \ket{x}\ ,
\end{align}
where $\sum_x |\alpha_x|^2 = 1$. We remark that this means that since there are
$2^n$ possible strings $x \in {\{0,1\}}^n$, we need an exponential number of
parameters $\alpha_x \in \Complex$ in order to describe the definite state
$\ket{\Psi}$. This is in sharp contrast to classical computing, where only $n$
parameters are needed (namely a specific string $x$).

As an example, if we have two qubits $A$ and $B$, and the first qubit is in a
state $\ket{0}_A$ and the second in a state $\ket{1}_B$, then the overall state
of the two qubits can be expressed as $\ket{01} = \ket{0}_A\ket{1}_B$.
However, there exists multi-qubit states $\ket{\Psi}$ which cannot be written
as such a combination of single qubit states.  That is, the two qubits can non
longer be described independently of each other.  The states of the two
individual qubits are now correlated beyond what is possible to achieve
classically.  Such states are called \emph{entangled}. For two-qubits the
maximally entangled state can (up to local quantum gates) be written as
\begin{align}
  \ket{\Phi} = \frac{1}{\sqrt{2}}\left(\ket{0}_A\ket{0}_B +
  \ket{1}_A\ket{1}_B\right)\ .
\end{align}
Such states have an interesting property that for any measurement on $A$ that
probabilistically yields outcome $x$, there always exists a measurement on $B$
that yields exactly the same outcome $x$. Very intuitively, such states can
hence be understood as the quantum analogue of maximal correlation in the
classical domain, only such correlations persist for any
measurement. Entanglement enables much stronger than classical correlations,
also for more complex scenarios~\cite{van2014quantum}.  Interestingly,
entanglement cannot be shared, which is also known as the monogamy of
entanglement~\cite{terhal2004entanglement}.

An entangled state is created from initially unentangled qubits, say
$\ket{0}_A\ket{0}_B$. A common scheme to locally create an entangled state is
to start by applying the so-called Hadamard operation on $A$ to produce
$(\ket{0}_A+\ket{1}_A) \ket{0}_B /\sqrt{2} $.  Subsequently a controlled NOT
operation (CNOT) is performed which has the effect $\text{CNOT}
\ket{x}_A\ket{y}_B = \ket{x}_A\ket{y+x \mod 2}$:
\begin{equation}
  \text{CNOT} \frac{1} {\sqrt{2}} \left( \ket{0}_A \ket{0}_B + \ket{1}_A
  \ket{0}_B \right) = \frac{1}{\sqrt{2}} \left( \ket{0}_A \ket{0}_B + \ket{1}_A
  \ket{1}_B \right).
\end{equation}
The physical implementation depends on the underlying hardware platform. For NV
centres in diamond this operation can be implemented using a combination of a
microwave and optical pulses~\cite{hensen2015loophole}.

\subsection{Teleportation}

\begin{figure}[h]
  \centering
  \includegraphics[width=\linewidth]{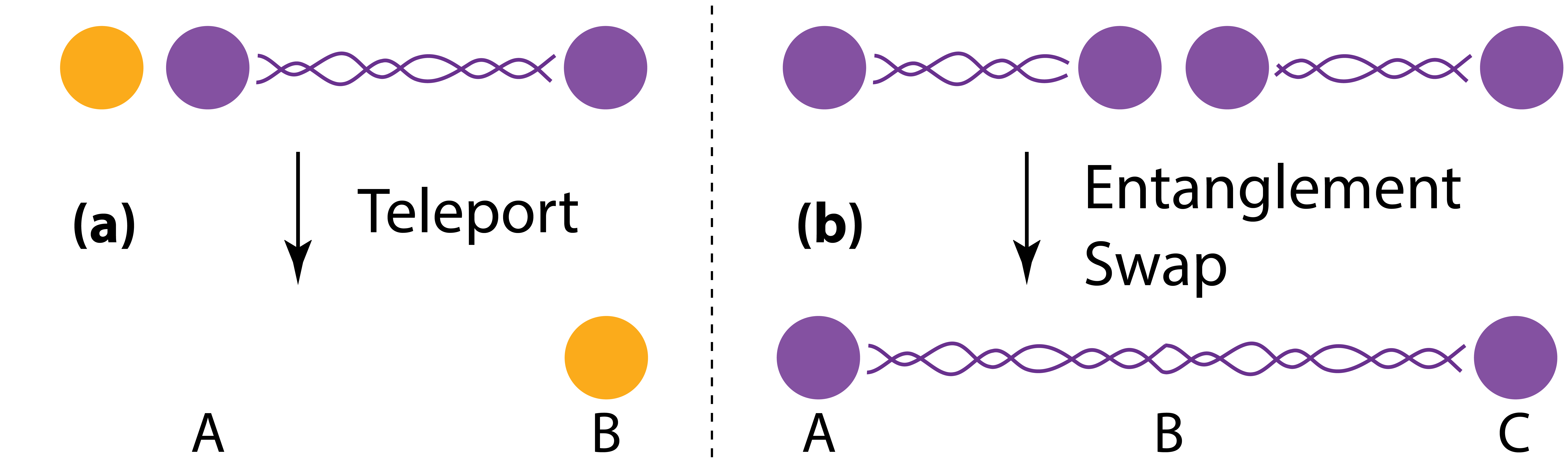}
  \caption{Entangled Bell pairs enable long-distance quantum
    communication. (a)~An unknown data qubit state can be teleported over long
    distances by consuming a Bell pair that has one qubit at the source and the
    other qubit at the destination. (b)~Two shorter Bell pairs can be combined
    into a longer Bell pair with an entanglement swap
    operation.}\label{fig:teleport_swap} \Description{Quantum teleportation and
    entanglement swapping.}
\end{figure}

Qubits may be transmitted directly, or via quantum
teleportation~\cite{bennett1993teleporting} using entanglement.  To teleport
one data qubit $\ket{\Psi}$, we require one entangled pair $\ket{\Phi}$ to be
established between the sender and receiver ahead of time.  The sender performs
a measurement of the data qubit $\ket{\Psi}$ and their qubit $A$ of
$\ket{\Phi}$ (see Fig.~\ref{fig:teleport_swap}), resulting in two classical
bits $y\in{\{0,1\}}^2$ as the measurement outcome. The sender transmits $y$ to
the receiver, who applies a correction depending on $y$ on their qubit in order
to recover $\ket{\Psi}$. From the perspective of control of such a network, we
remark that this requires that the sender has correctly identified that qubit
$A$ belongs to the entangled state $\ket{\Phi}$ shared with the receiver, and
that the entanglement is consumed by this process. Deterministic teleportation
has been realised, using for example two network nodes based on NV in
diamond~\cite{pfaff2014unconditional}.

Teleportation is crucial for quantum networking. The no-cloning theorem means
that retransmitting the data qubit if sending fails is not an option. However,
$\ket{\Phi}$ is a known generic state that does not carry any data and can be
repeatedly recreated until it has been successfully distributed to the sender
and receiver. At this point the sender simply teleports the sensitive data
qubit to the receiver without putting it through the network risking its loss.

\subsection{Entanglement Swapping}

Teleportation also provides a mechanism to extend short-distance entanglement
to larger distances~\cite{munro2015inside, briegel1998quantum, dur1999quantum}.
Consider node $A$ which has generated entanglement with node $B$. Similarly,
$B$ has produced entanglement with $C$. We can now generate entanglement
between $A$ and $C$ using the help of $B$: $B$ teleports the qubit entangled
with node $A$ to $C$, using the entanglement he shares with $C$. This process
is also known as entanglement swapping~\cite{briegel1998quantum,
  zukowski1993event} (see Fig.~\ref{fig:teleport_swap}).

Unfortunately, neither the entanglement generation nor the swapping operations
are noiseless.  Therefore, with each link and each swap the quality of the
entanglement, called fidelity, degrades. However, it is possible to create
higher fidelity entangled pairs from two or more lower quality pair states
through a process called distillation using the Purify-and-Swap
algorithm~\cite{briegel1998quantum}. Therefore, once the quality loss over a
given distance become prohibitive, additional redundancy may be used to restore
the state fidelity.

\section{Elements of a Quantum Network}\label{sec:elements-of-a-quantum-network}

Let us provide a high-level overview of the elements of a quantum
network~\cite{wehner2018quantum}.  For additional overview of design
considerations for quantum networks we also refer to
Refs.~\cite{dahlberg2019link, van2014quantum, van2013designing}.

\paragraph{End Nodes:}
Just like in classical networks we need devices at the edge of the network on
which applications are run. In the simplest case, these are photonic devices
consisting of linear optical elements, photon sources and detectors. These do
not have a quantum memory to store qubits, and can also only perform a limited
set of quantum operations deterministically. However, these are sufficient to
perform all protocols in the prepare and measure stage of quantum
network~\cite{wehner2018quantum} at short distances (presently
\textasciitilde{}100~km over deployed telecom fibre), such as QKD.\@

However, they may also be \emph{processing nodes} with an optical interface
which are capable of storing qubits, as well as performing universal quantum
computation.  Examples include NV centres in diamond~\cite{hensen2015loophole,
  bernien2013heralded, taminiau2014universal}, ion
traps~\cite{moehring2007entanglement}, and neutral
atoms~\cite{reiserer2015cavity}. Such systems can also be used to run
application protocols in the quantum memory network stage and eventually
above~\cite{wehner2018quantum}.

\paragraph{Quantum repeaters:}
The objective of quantum repeaters is to transmit qubits over
long-distances. Any system that is a quantum processing node, can also be used
as a repeater platform. In addition, there exist specific hardware platforms
tailored to the task of a quantum repeater. This includes multiplexed quantum
repeaters~\cite{sangouard2011quantum} which promise to generate entanglement
quickly by temporal and spatial multiplexing. These repeater platforms work ---
in a possible combination with entanglement distillation steps --- by the
entanglement swapping principle outlined in
Fig.~\ref{fig:teleport_swap}. Theoretical proposals for employing forward error
correction also exist~\cite{muralidharan2014ultrafast}, but they are not
possible to realise in the near-term.

\begin{figure}[h]
  \centering
  \includegraphics[width=\linewidth]{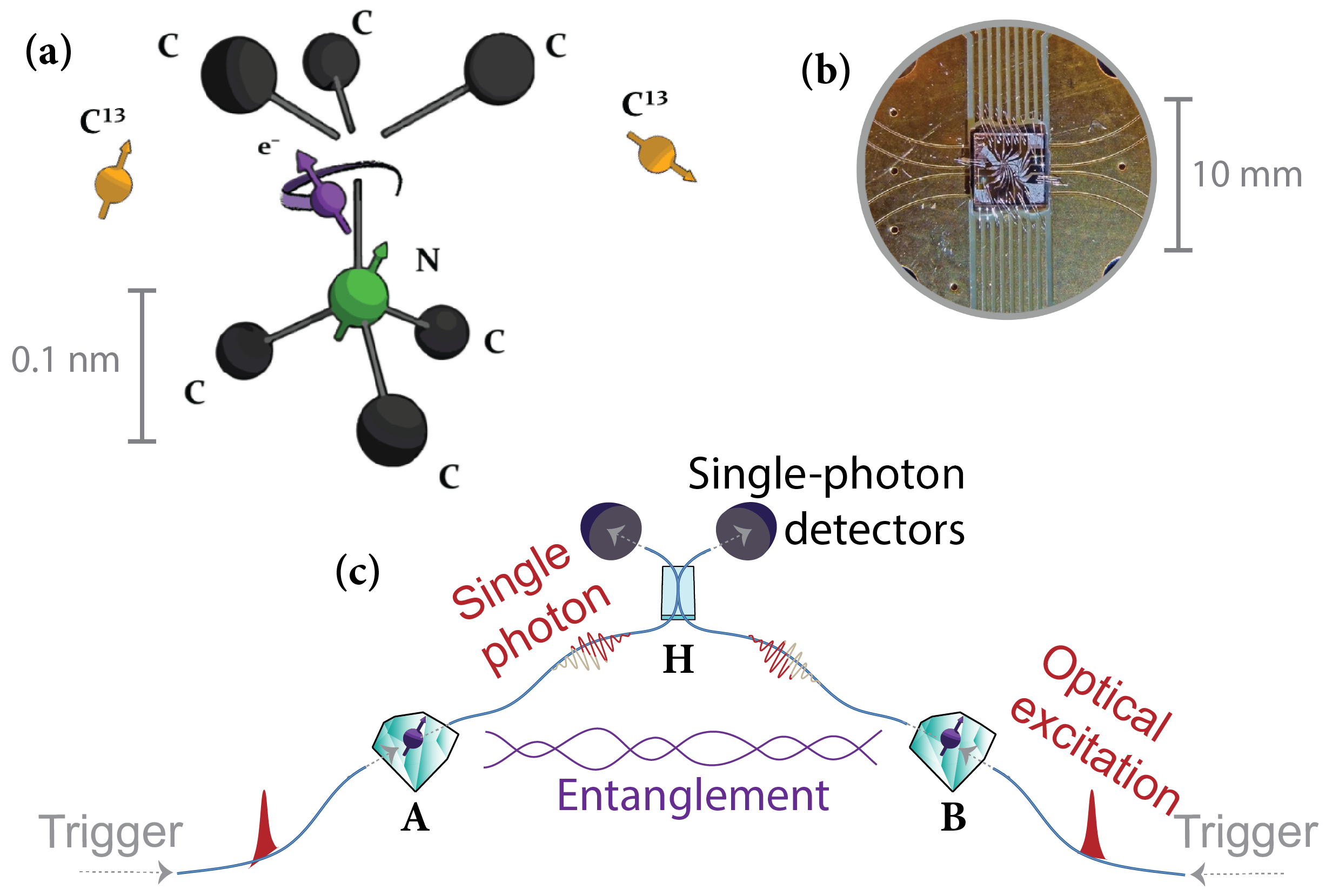}
  \caption{Example of a physical implementation producing entanglement between
    two quantum processors (NV in diamond).  (a) The qubits of the Bell pair
    are stored in NV centres in diamond on (b) custom chips. (c) Entanglement
    is generated between the two processors using probabilistic entanglement
    swapping: entanglement is produced between each processor and a traveling
    qubit (photon) sent to the mid-point. The mid-point performs entanglement
    swapping, and sends a confirmation (heralding) signal back to the nodes
    whether entanglement generation was a success.}\label{fig:experiment}
  \Description{Quantum link experimental setup.}
\end{figure}

The current world record for producing such heralded (i.e.\ confirmed)
entanglement is 1.3~km which has been achieved using NV centres in
diamond~\cite{hensen2015loophole}, see Fig.~\ref{fig:experiment}.  This
platform is a few (about 10~\cite{bradley201910}) qubit quantum computer with
an optical interface capable of executing arbitrary gates and measurements.  It
has been recently demonstrated that NV centres are capable of memory lifetimes
approaching one minute~\cite{bradley201910} in nodes not yet interfaced to the
network.  Other platforms exist that are similar on the conceptual level with
similar capabilities such as ion traps~\cite{inlek2017multispecies} and neutral
atoms~\cite{reiserer2015cavity} (see Table~\ref{tab:QLE} for current parameter
trade-offs).

\paragraph{Communication lines:}
Qubits can be sent using photons through fibre, or free space
communication~\cite{yu2019entanglement}. Standard telecom fibre can be used for
this purpose, potentially following an appropriate wavelength conversion to the
telecom band~\cite{dreau2018quantum, zaske2012visible}.

\begin{table}
  \caption{Quantum Link Efficiency (QLE) is given by the ratio of the
    entangling rate to the decoherence rate, capturing how fast entanglement
    can be produced in relation to how fast it is lost. A QLE $\geq 1$ is
    required to extend entanglement over long distances.}\label{tab:QLE}
  \begin{tabular}{c c c c}
    \toprule
    Platform & QLE \\
    \midrule
    NV Centres & 8~\cite{humphreys2018deterministic} \\
    Trapped Ions & 5~\cite{hucul2015modular} \\
    Neutral Atoms & 2 (projected)~\cite{nolleke2013efficient,
      korber2018decoherence} \\
  \bottomrule
\end{tabular}
\end{table}

\paragraph{Classical Control Messages}
A crucial component of quantum communication is also the ability to send
classical data. The control of quantum devices requires quite a number of
classical control signals to be exchange, teleportation being just one
example. In order to develop functional quantum protocols we will need a way to
transmit control information between the quantum repeaters.  This means that it
is expected that quantum networks will be deployed alongside classical networks
with a quantum data plane coexisting with the classical one as shown in
Fig.~\ref{fig:network}.

\section{A Quantum Network Stack}\label{sec:a-quantum-network-stack}

\begin{figure}[h]
  \centering
  \includegraphics[width=\linewidth]{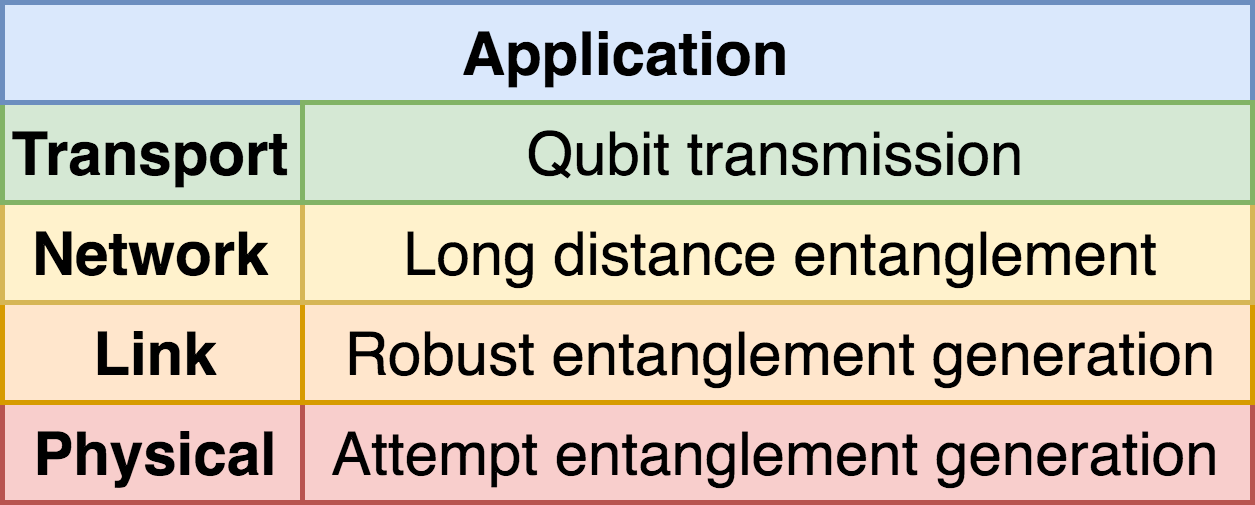}
  \caption{Functional allocation in a quantum network stack. The structure
    mirrors and is inspired by the classical TCP/IP network
    stack.}\label{fig:stack} \Description{Quantum network stack.}
\end{figure}

One may wonder whether one can design quantum network protocols without
detailed knowledge of the underlying hardware system.  Here, we briefly
summarise the approach of Ref.~\cite{dahlberg2019link}, because it is defined
in terms of service layers rather than protocol layers (see
Fig.~\ref{fig:stack}) which gives it a structure that is similar to the
classical TCP/IP stack. It also gives a concrete link layer protocol that
abstracts away from the underlying hardware system, turning entanglement
generation into a well-defined service.

\paragraph{Physical}
This layer corresponds to the actual quantum hardware devices and physical
connections. The physical layer keeps no state related to entanglement
production, produced entanglement probabilistically, and has no decision making
capabilities. The hardware is solely responsible for tasks such as time
synchronisation, photon emission, laser phase stabilisation, and so on, that
are required to actually produce entangled Bell pairs.

\paragraph{Link}
The task of the link-layer is to utilise the physical layer's ability to
produce entanglement between neighbouring nodes reliably. It also integrates
the quantum and classical data planes providing sufficient information for
higher level protocols and network management.  A concrete link layer protocol
can be found in~\cite{dahlberg2019link}.

\paragraph{Network}
Similar to a network layer in classical networking, the task of the network
layer is to enable the generation of entanglement between network nodes which
are not directly connected. A protocol to achieve this would utilise the link
layer to produce entanglement between neighbouring nodes followed by
entanglement swaps to create long distance links.

\paragraph{Transport}
One can imagine, that a transport layer could provide the additional service of
transmitting qubits to the application layer.  This could be realised by, for
example, pre-generating entangled pairs of qubits using the network layer,
followed by teleportation to ensure reliable end-to-end delivery of qubits.

\section{Challenges and Requirements}\label{sec:challenges-and-requirements}

Quantum networks are still in their early infancy. Realising the necessary
quantum hardware is of paramount importance and presents many challenges, but
that is only one part of the story.  Here, we present some of the challenges
beyond hardware accounting for the fundamental differences inherent to quantum
communication and mitigating the limitations and imperfections at the physical
level.  Further design considerations can also be found
in~\cite{dahlberg2019link}.

\paragraph{Timely decision making}
Quantum memory lifetimes are extremely short even in the most sophisticated
setups and this directly impacts our ability to produce long-distance
entanglement by means of entanglement swapping. Entanglement swapping requires
that both entangled pairs of qubits are available on two separate links at the
same time so the intermediate node must be able to store the first pair until
it receives the second pair. If one of the qubits decoheres, the pair is lost
and the entire process must start over. One approach to increase the likelihood
of such a coincidence event lies in proposals to perform massive
multiplexing~\cite{sangouard2011quantum} significantly reducing the required
storage time. There is also the obvious approach of increasing memory
lifetime. NV centres in diamond already exhibit a high QLE, see
Table~\ref{tab:QLE}, and lifetimes up to a minute have recently been observed
in NV nodes not yet connected to a network~\cite{bradley201910}. Longer memory
lifetimes impose less stringent demands on timing at the network layer allowing
it to be kept at the physical layer.

Nevertheless, mitigating limited qubit lifetimes is essential and demands fast
and reactive control of the network. In a network based on entanglement
swapping it also raises the interesting question of whether such entanglement
is produced only on-demand, or if there exists a mechanism which continuously
generates entangled pairs at all times between certain links of the network.

\paragraph{Extending the network stack}
In parallel with the effort of building the physical network links there is a
need for work to build up the quantum network stack vertically. The first
link-layer protocol has been proposed~\cite{dahlberg2019link}. However, to go
beyond point-to-point connectivity between two directly connected nodes we need
a network layer service and the transport layer to provide platform-independent
services for distributed quantum applications.  The first end-to-end quantum
communication protocols have started to appear though they generally assume
hardware capabilities beyond what is possible in the near-term
future~\cite{matsuo2019quantum, yu2019protocols}.

\paragraph{Routing}
In addition to forwarding protocols necessary to actually generate an
end-to-end Bell pair there are many other second-level mechanisms necessary for
a fully functional quantum internet. The specific question of routing
entanglement, i.e.\ making decisions on how end-to-end entanglement can be
established quickly between users in future quantum networks, is seeing more
attention~\cite{caleffi2017optimal, gyongyosi2018decentralized, van2013path,
  schoute2016shortcuts, gyongyosi2017entanglement, pant2019routing}.  Routing
in quantum networks is a non-trivial problem due to the non-local and temporary
nature of entangled pairs as well as different physical resource requirements
necessary for delivering these pairs with a high enough fidelity.

\paragraph{SDN Integration}
Given limited lifetimes, building robust and efficient quantum network routing
and management protocols in an entirely distributed manner may be difficult.
This could make software-defined networking (SDN) a very attractive direction
for quantum networking and has already been considered for
QKD~\cite{ou2018field}.

SDN is an architecture for programmable networks that splits the vertical
integration of the forwarding and control planes and puts much of the
decision-making capabilities in a centralised controller (physically
decentralised with appropriate redundancy)~\cite{kreutz2015software}.  In this
approach, the central controller has network-wide visibility and it is
responsible for most (or all as is the case for OpenFlow) control plane
decisions based on input it receives from the individual nodes in the network.
It is plausible that in a quantum network a controller would be responsible for
managing the global strategies for the distribution of long-distance Bell pairs
(Bell pairs that have been produced as a result of entanglement swaps between
separate links), but connection establishment, Bell pair generation, and other
localised operations are left to the actual devices who will try to conform to
the controller's strategy.

\paragraph{Security}
Given that one of the most important features quantum networking brings with it
is enhanced security it is crucial that a design for a future quantum network
architecture incorporates strong security features itself. Such design
considerations should be employed already at the physical layer, to ensure the
protection of quantum network nodes. For example, we remark that convincing a
remote node to produce entanglement with its neighbour may simply lead to a
denial of service attack consuming its resources~\cite{dahlberg2019link}. This
shows that at the very least authentication is necessary for control messages
already at the physical layer. Such authentication could be realised using
standard classical mechanisms, or also use keys generated by QKD in combination
with an information-theoretically secure authentication scheme.

\section{Conclusion}\label{sec:conclusion}
There is a tremendous amount of work to do to build a fully functional quantum
network, both at the physical level and at the systems and software
level. Recent experimental progress in entanglement generation rates and memory
lifetimes is very promising and the breadth of the combined research effort
should result in practical demonstrations very soon. Nevertheless, there are a
lot of open questions and research challenges that are unresolved and require a
range of expertise from beyond physics such as operating systems, computer
networks, and communications. This opens up many new opportunities for
researchers from outside the usual circles to contribute to the growing field
of quantum networking.

\section*{Acknowledgements}

The authors of this memo acknowledge funding received the EU Flagship on
Quantum Technologies, Quantum Internet Alliance, an ERC Starting Grant (SW) and
an NWO VIDI Grant (SW).  The authors would further like to
thank~\cite{dahlberg2019link} for permission to reuse some of their figures.

\bibliographystyle{ACM-Reference-Format}
\bibliography{bibliography}

\end{document}